# Spatial frequencies associated with the latitudinal structures of ionospheric currents seen by CHAMP satellite

**Neethal Thomas, Geeta Vichare[1], A. K. Sinha**

Indian Institute of Geomagnetism,

Plot 5, Sector 18, New Panvel (W), Navi Mumbai- 410218, India

**Corresponding Author:** Neethal Thomas (neethalmariyathomas@gmail.com)

[1]*Formerly Geeta Jadhav*

**Abstract**

The CHAMP magnetic field variations during international quiet days of low solar activity period 2008-2009 are investigated. The present paper reports the existence of frequency-peaks ≤20 mHz in the compressional component of the magnetic field in almost all CHAMP passes. The magnetic field variations associated with these frequencies have amplitude of a few tens of nT during daytime. The geomagnetic activity and interplanetary magnetic field parameters were observed to be low during the period of study. The spectral powers of the observed frequencies show no dependence on solar wind velocity and cone angle; hence the reported frequencies are not related to the geomagnetic pulsations. For frequency-peaks ≤15 mHz, strong local-time dependence is observed with maximum power near noon and minimum at night. The longitudinal and seasonal variations of the powers of these frequency-peaks match well with those of the equator-to-middle latitude ionospheric currents derived by the earlier studies. As a polar Low-Earth-Orbiting (LEO) satellite spans the entire range of latitudes within few minutes, it monitors the geomagnetic field variations caused by the quiet-time ionospheric currents flowing at different latitudes. This can result in certain frequencies in the magnetic field recorded by LEO satellites. We demonstrate that the frequencies <10mHz are mainly due to the latitudinal structure of the equatorial electrojet. The observed frequencies in CHAMP data are therefore attributed to the latitudinal structures of the ionospheric currents that are monitored only by the polar LEO satellites and are found to alter the observations of geomagnetic pulsations (Pc4-5 and Pi2) significantly.

1. Introduction

The magnetic field measurements from Polar Low Earth Orbiting (LEO) satellites are widely used to study various ionospheric and magnetospheric processes such as ionospheric E-region currents, F-region currents, equatorial plasma bubbles (EPBs), spread-F, geomagnetic pulsations etc. [*Jadhav et. al.,* 2001, 2002; *Lühr et. al.,* 2004, 2015; *Stole et. al.,* 2006; *Park et. al.,* 2009, 2013; *Pedatella at. al.,* 2011; *Vellante et. al.,* 2004; *Ndiitwani et. al.,* 2009]. Due to the motion of the polar LEO satellite, it can also observe the spatial structures of these processes [*Stole et. al.,* 2006; *Park et. al.,* 2009; *Lühr et. al.,* 2014; *Nakanishi et. al.,* 2014; *Iyemori et. al.,* 2014;]. This can result in the observation of certain frequencies in the magnetic field recorded by LEO satellites. Therefore, the magnetic field variations observed by the LEO satellite can have varied frequencies, which can be of temporal (geomagnetic pulsations) or spatial (due to the motion of the satellite) origin.

The spatial structures of various quiet time ionospheric phenomena with different periodicities have been noticed by previous researchers in the magnetic field measurements at LEO satellites. While studying night time F-region currents using CHAMP total magnetic field measurements, *Luhr et. al.,* [2002] observed a close association of these currents with small spatial scale (~100 Km) intensity fluctuations, which they attributed to F-region irregularities like spread F phenomenon. Later, using vector magnetic field measurement from CHAMP, *Stole et. al.,* [2006] studied the magnetic deflections associated with post sunset equatorial spread F (ESF) in detail. They could observe the finest spatial scales of ESF to around few tens of meters, resulting in a periodicity <30s in the parallel magnetic field component at CHAMP. Similar observations of magnetic fluctuations associated with equatorial plasma bubbles (EPBs) were deduced by *Park et. al.,* [2009a] in the CHAMP data. *Park et. al.,* [2009b] reported the occurrence of spatial structures having periodicities <30 s in the zonal and meridional components at CHAMP during night hours, which they attributed to mid latitude magnetic fluctuations (MMFs). The characteristics of observed MMFs were found to be consistent with medium scale travelling ionospheric disturbances (MSTID) and are attributed to spatial structures of FACs. In a comprehensive survey of electromagnetic signatures related to EPBs, *Luhr et. al.,* [2014] observed magnetic field fluctuations in the transverse components having periodicities <1 s in the post sunset hours, which were interpreted as Alfvénic signatures (carried by FACs) accompanying plasma density depletions of scales lengths 150m–4km. *Nakanishi et. al.,* [2014] reported the presence of global and frequent appearance of small amplitude fluctuations with 10-30 s period in the zonal and meridional components of the magnetic field measurements at CHAMP. Similar observation was also confirmed by *Iyemori et. al.,* [2015] using SWARM mission satellites. These oscillations from mid-to-low latitudes showed strong LT and seasonal dependence; and were attributed to the spatial structures of FACs generated by the ionospheric dynamo driven by the atmospheric gravity waves propagating from the lower atmosphere to ionosphere. Thus in the low-to-mid latitudes, the magnetic field fluctuations in various components having periodicities <30s (frequency >33 mHz) associated with the spatial structures of ionospheric irregularities and FACs monitored by the polar LEO satellites are well noticed and accounted in the literature.

The ultra low frequency (ULF) oscillations of earth's magnetic field called as geomagnetic pulsations are recorded as temporal oscillations in the magnetic field measurements. The geomagnetic pulsations, which are the manifestation of magnetohydrodynamic (MHD) waves, are produced by various processes in the earth's magnetosphere and solar wind. Therefore, the study of pulsations is of great significance as it can serve as a diagnostic tool for the identification of space weather events. Furthermore, the observations of these ULF oscillations using LEO satellites stand advantageous as it provides an opportunity to study the geomagnetic pulsations from the topside of the ionosphere and hence to investigate the effects of the ionosphere on the propagation of geomagnetic pulsations.

The continuous pulsations (Pcs) in Pc1-4 bands (frequencies >15 mHz) are analyzed using LEO satellite magnetic field measurements [*Jadhav et al.,* 2001; *Vellante et. al.,* 2004; *Heilig et. al.,* 2007; *Pilipenko et. al.,* 2008; *Ndiitwani et. al.,* 2009; *Park et. al.,* 2013; *Sutcliffe et. al.,* 2013; *Cuturrufo et. al.,* 2014; *Yagova et. al.,* 2015]. However, the observation of geomagnetic pulsations with frequencies <15 mHz (comprise Pc5 and part of Pc4 bands) are mostly confined to magnetospheric satellites and ground, and are not well reported at LEO. Similarly, the impulsive geomagnetic pulsations with frequency range between 6.6-25 mHz, termed as Pi2s are well studied using LEO observations during nighttimes [*Takahashi et. al.,* 1999; *Han et. al., 2004*; *Sutcliffe and Luhr* 2010; *Cuturrufo et. al.,* 2014; *Thomas et. al.,* 2015], nevertheless the daytime observation of Pi2s at LEO revealed different conclusions about its detection at topside ionosphere and hence remained puzzling [*Han et. al.,* 2004; *Sutcliffe and Luhr* 2010]. In a recent article, *Thomas et. al.,* [2015] showed that daytime compressional Pi2s at CHAMP can be severely contaminated by the background low frequencies (<15 mHz) those are particularly present in satellite observations. They attributed those background frequencies to the spatial structures of E-region ionospheric currents. They also demonstrated that upon eliminating these frequencies, clear daytime Pi2s can be observed at CHAMP.

While studying Pc3 pulsations, *Heilig et. al.,* [2007] noticed the presence of lower frequency (<10 mHZ) in the compressional component at CHAMP (their Figure 5). Although not examined in detail, they attributed its origin to the inadequate removal of the crustal anomalies by the ambient magnetic field model. Thus, the background lower frequencies observed by *Heilig et. al.,* [2007] and *Thomas et. al.,* [2015] were believed to be of non-temporal origin. In a comparative study of ULF oscillations in Pc3-5 bands (1-100 mHz) using CHAMP and two magnetospheric satellite missions (Cluster and Geotail), *Balasis et. al.,* [2012] reported that the Pc4-5 waves could not be detected at CHAMP, whereas magnetospheric satellites could detect the pulsations from all the three bands of Pc3-5. They attributed these observations to the fast motion of CHAMP satellite through field lines at lower altitudes. Also *Thomas et. al.,* [2015] suggested that the Pi2s in the compressional component from the higher frequency range (>15 mHz) are more suitable for the detection in the topside ionosphere using polar LEO satellites. Thus in general, the observation of temporal oscillations (Pis or Pcs), at LEO are mainly limited to higher frequencies (>15 mHz). In this context, the observation of background lower frequencies (<15 mHz) as reported by *Heilig et. al.,* [2007] and *Thomas et. al.,* [2015] at LEO satellites may provide some insight into the possible reasons for the non-detection of geomagnetic pulsations with lower frequencies at LEO.

Although these lower frequencies are noticed at CHAMP by *Heilig et. al.,* [2007] and *Thomas et. al.,* [2015], they related its origin to different mechanisms. *Thomas et. al.,* [2015] ascribed these frequencies to the spatial structures of the ionospheric current systems monitored by the LEO satellite in its transit from pole to pole. But, *Heilig et. al.,* [2007] attributed to the crustal anomalies present in the satellite measurements. Hence, there are different opinions about the origin of these frequencies. Also as described above, the magnetic field variations of spatial origin related to various ionospheric phenomenon are only reported for frequencies >33 mHz, by previous researchers. Therefore the spatial structures having frequencies <30 mHz are less explored. In these perspectives, detailed investigation of these frequencies is important. Firstly, we establish the existence of the frequency peaks in the total magnetic field variations (compressional) at the satellite, during geomagnetic quiet conditions. Here, we detail for the first time the origin of these frequencies by investigating their characteristics.

The paper is organised as follows: Section 2 presents the data set and the method of identifying frequency peaks. Section 3 examines the solar wind (SW) parameters prevailed during the period of the present study, to estimate the level of contribution of geomagnetic pulsations in the identified frequency peaks. Section 4 demonstrates the effect of these frequencies on the temporal oscillations of Pc4-5 waves. The local time (LT), longitudinal and seasonal variations of the observed frequency peaks are presented in Section 5. Section 6 discusses and summarizes the findings.

## 2. Data and method of analysis

Magnetic field measurements obtained from the CHAMP satellite during international quiet days of two years (2008-2009) are utilized in this study. The solar minimum period considered for the present analysis ensures the low level of geomagnetic activity. The German satellite CHAMP was launched on July 2000 into a near circular polar orbit and lasted for 10 years in orbit till September 2010. In this duration, the CHAMP descended in altitude from initial 456-250 Km. With an orbital period of ~92 min, CHAMP orbits around the Earth 15 times a day constituting ~15 day and night passes each. Its orbit drifted slowly in LT descending 1 h in 11 days and hence monitored all LT sectors in approximately 4 months duration. The considered two years of measurements therefore, give good global coverage of the magnetic field observations over all LT sectors.

The magnetic field data used in the present study are obtained from the fluxgate magnetometer onboard CHAMP with 1s sampling rate. As magnetic field measured at CHAMP has contributions from various geomagnetic sources of internal and external origin, we have subtracted the geomagnetic field model (Potsdam Magnetic Model of Earth) POMME 6.1 [*Maus et. al.,* 2010] from the magnetic field recorded by CHAMP. POMME model accounts for the main geomagnetic field and its secular variation and acceleration. It also provides contributions from static magnetospheric currents and the induction effect caused by the time varying ring current. Therefore, the residual field obtained after subtracting POMME has contributions from various quiet time ionospheric currents such as EEJ, Sq, FAC etc. It can also have contributions due to disturbed time ionospheric and magnetospheric processes. Since the period of present work is geomagnetically quiet, the disturbed time contributions can be minimal. For this study, we have utilized 120 days comprising first five international quiet days of each month of the years 2008-2009, which provided total ~ 3600 (30x120) satellite passes, constituting 1800 day and night passes each. Only CHAMP passes within ±50° magnetic latitude are considered for the study to avoid complex and intense high latitude contribution. In order to remove the broad trend, a high pass filter with cut off frequency of 1.5 mHz is applied to the residual total magnetic fields. Here, it should be noted that the residuals in the total magnetic field is the same as that in the compressional component.

The PSD of the high pass filtered total magnetic field is estimated for each pass, using Maximum Entropy Method (MEM). MEM spectrum is computed for a data window of 1600 points with an Auto Regression order (length of prediction filter) 300. A typical profile of the residual total magnetic field observed by CHAMP (red curve) during daytime between ±50° magnetic latitudes is shown in Figure 1a. Figure 1b shows the normalised (by variance) PSD of the observed residual total magnetic field (red curve) along with 99% confidence level. Considering the MEM spectrum to be Chi-square distributed with two degrees of freedom, the confidence level equals to 99[th] percentile value for Chi-square times the normalized red noise spectrum [*Gilman et. al.,* 1963]. The frequency peaks identified between 3-21 mHz are considered in the study. Prominent frequency peaks (at ~ 4.5, 7, 9, 13, 17 and 21 mHz) are marked by red arrows. Although, a few satellite passes showed frequency peaks >21 mHz, they were not found consistent and hence not considered in the present study.

Since we have removed the background ambient magnetic field from CHAMP observations, the residual field depicted in Figure 1a represents the magnetic field variations due to external sources like quiet time ionospheric currents. The negative depression near dip equator and the positive lobes on its either side clearly mark EEJ and the associated return current signatures respectively in the total magnetic field variations [*Jadhav et. al.,* 2002 and *Luhr et. al.,* 2004]. Also, the profile displayed in Figure 1a shows slight dip between ~25°-35° magnetic latitude in both the hemispheres, which could be due to superimposed effect of Sq and EEJ return currents.

Since Polar LEO satellite traverses over the entire latitudinal range in short time (~45 min), it predominantly records the latitudinal (spatial) variations rather than the temporal variations of the ionosheric currents. As a result the spatial structures of varied wavelengths associated with the latitudinal profile of ionospheric currents can be present in the magnetic field measurements at the satellite. The orbital period being ~92 min, CHAMP covers ~4° of latitude in one minute. The frequency of 3 mHz, corresponds to time period of ~5.5 min. In this duration the satellite covers ~22° of latitudinal extent, which is referred to as latitudinal wavelength. Thus the present study examines the spatial structures having latitudinal wavelengths <22°.

A typical signature of EEJ has amplitude over 20 nT and latitudinal width between ~8°-12°, which would be covered by CHAMP in ~2-3 min. Therefore, the latitudinal profile of EEJ with the periodicity of ~2-3 min would result in a frequency of 8-6 mHz. Although, it should be noted that the width and amplitude of EEJ has strong day-to-day and longitudinal variability [*Jadhav et al.* 2002]. ]. The day to day variability in the EEJ current pattern is strongly associated with the variability in the thermospheric neutral tidal winds. Since the typical EEJ signature being encompassed of a negative depression and two positive lobes of variable amplitudes, it can give rise to multiple frequency peaks of variable power. In order to demonstrate this point, we have plotted the EEJ profile obtained by fitting empirical model given by Onwumechili [1997] to the CHAMP observations (Blue curve in Figure 1a). The PSD of the fitted EEJ profile is shown in Figure 1c, which shows prominent frequency peaks at ~4.5 mHz, 7 mHz and 9 mHz (marked with blue arrows), demonstrating the existence of different frequencies corresponding to the displayed EEJ profile in Figure 1a. Interestingly, the frequency peaks of the fitted EEJ profile (Figure 1c) coincide well with the first three frequency peaks of the observed residual field (Figure 1b), with the identical powers. Nevertheless, it should be noted that the number of frequency peaks and their centre frequencies varies with the shape and size of the EEJ profile. Normally only two peaks are observed in the fitted EEJ profile, which match with the first two frequencies of the observed profile. This may indicate that first two to three peaks (<10 mHz) in the observed residual field are associated with the latitudinal structure of EEJ, as powers of the coinciding frequencies are nearly the same. It is possible that similar wavelengths can be associated with the latitudinal structures of other current systems, but EEJ being the strongest among the currents flowing in the equator to mid latitude region, the powers of frequencies associated with other ionospheric currents can be of smaller magnitudes. The observed higher frequency peaks may be associated with the smaller scale latitudinal wavelengths of other currents systems such as ionospheric dynamo currents driven by gravity waves, Sq etc.

Frequency peaks and the related power are identified for all the selected 3600 passes using an automated program. The program identifies the frequency peaks >3 mHz which are above 99% confidence level. The frequencies identified are then classified into six bins between 3-21 mHz, each of width 3 mHz. As discussed before, the observed frequencies can be attributed to the spatial structures of various daytime ionospheric currents, observed by spacecraft at LEO. Different frequency bins and equivalent approximate latitudinal wavelengths are shown in Table 1. If no well defined frequency peak is identified in a bin, then it is marked as 'absent'. The frequency peaks beyond 21 mHz are not considered in the study because their power is very small and the occurrence is rare. In the present study, significant frequency peaks (>99% confidence level) between 3-21 mHz were evident consistently in almost all the satellite passes.

## 3. Geomagnetic activity and solar wind parameters

Since the observed frequencies (3-21 mHz) in the present analysis overlap with the Pc4-5 (2-22 mHz) and Pi2 (6-25 mHz) pulsations, it is of primary importance to check the contributions of the temporal oscillations in the frequencies reported here.

It has been known since long time that the continuous geomagnetic pulsation activity in Pc3-5 (~2-100 mHz) frequency band is well correlated with solar wind (SW) parameters [*Saito* 1964]. Dayside Pc3-4 pulsations observed mainly at low-middle latitudes were reported to have its origin in the upstream solar wind. The amplitude, power and the occurrence rate of these pulsations are reported to have good correlation with SW speed, interplanetary magnetic field (IMF) strength and cone angle [*Bol'shakova and Troitskaya* 1968; *Singer et al.,* 1977; *Wolfe et. al.,* 1980; 1985]. The cone angle, which is the angle between the sun-Earth line and the direction of IMF, is a key factor governing the occurrence of daytime Pc3-4 pulsations. A low cone angle (<30°) is considered to be the favourable condition for the generation of dayside Pc3-4 pulsations [*Greenstadt and Olson* 1976; *Russell et al.,* 1983; *Yumoto et al.,* 1985]. Pi2 pulsations are primarily associated with substorm activity [*Olsen* 1999], and hence the level of geomagnetic activity mainly decides the occurrence of Pi2s.

Therefore, in order to check the contribution of the geomagnetic pulsations in the observed frequencies, we firstly examine the geomagnetic activity and the solar wind conditions prevailed during the considered period of study.

The histograms of $\sum Kp$ index, SW and IMF parameters during considered 120 quiet days are shown in Figure 2. The $\sum Kp$ is found to be $\leq 5$ for ~95% of the days denoting the fairly good quiet geomagnetic conditions (Figure 2a). The average values of the SW and IMF parameters are estimated during each satellite pass (~25 min). The average proton density (Np) is found to be in general low with ~95% cases having $Np \leq 10$ (Figure 2b). The average SW velocity was mostly between 250-400 Km/sec and on a few occasions (< 15%) SW velocity was between 400-500 km/s (Figure 2c). The cone angle (Figure 2d) is found to be >30° for more than 85% of satellite passes. It is also to be noted that the magnitude of total IMF (Figures 2e) was very small mainly around 2 nT and the IMF Bz (Figures 2f) was mostly near zero level or northward. Thus the SW parameters depicted here ensure low solar activity existed during the considered days of study.

If the frequencies observed in the present study are of temporal origin, then one would expect a clear dependence of the power of the observed frequencies on SW velocity and cone angle [*Arthur and McPherron* 1977; *Takahashi et al.,*1981; *Heilig et. al.,* 2007] and are plotted in Figures 3 and 4. We have computed the average variation of the powers in each frequency bin with SW velocity by taking a running average of PSD for a data window of 20 Km/s. The error bars (vertical lines) represent the standard error. Similarly, the cone angel dependence of the average power in each frequency bin is computed for a data window of 10° stepping it forward by 5°.

It is interesting to note that the average power did not show any linear dependence on both SW velocity (Figure 3) and cone angle (Figure 4). These observations indicate that the frequencies observed in the present analysis are not of SW origin. *Heilig et. al.,* [2010] have reported that the Pc3-4 activity completely pauses during the periods of extremely low Np. Therefore, we have analyzed the period when Np was ~0.9 cm$^{-3}$. The other SW parameters prevailed during this selected period were: Kp =2-, SW velocity=~470 Km/s, cone angle=~80° and IMF Bz ~=1 nT. The PSD of the total magnetic field during the considered CHAMP pass is shown in Figure 5. Here, it is observed that even with this low Np level (Np <1), multiple frequency peaks from different frequency bins are still evident in the CHAMP observations with significantly high powers. These observations indicate that, frequencies observed in the total magnetic field measurements at CHAMP during its quiet time passes between ±50° magnetic latitude, are not associated with pulsations.

## 4. Modification of Pc4-5 oscillations at CHAMP

Recently, *Thomas et. al.,* [2015] reported that the Pi2 oscillations are highly contaminated during day hours by the background lower frequencies (<15 mHz) present at CHAMP. In this section we present a case study showing the possible interference of these background frequencies on the signatures of Pc4-5 pulsations observed by the CHAMP satellite. The occurrence of continuous Pc5 activity during the quiet period of 1-8 August 2008 prior to a CIR-induced storm on 9 August 2008 was reported by *Urban et. al.,* [2011]. Total magnetic field measurements from a satellite pass during this period of Pc5 activity is presented here to depict the manifestation of Pc4-5 signatures at CHAMP.

Figure 6 shows the Pc4-5 event observed at ground station BOU (Boulder: Geographic Coordinates: 40.1°N, 254.8°E) and the simultaneous observations from over head CHAMP during daytime, on 01 August 2008 between 1515-1536 UT. Figure 6a shows the PSD of the satellite compressional and BOU H components (high pass filtered above 0.5 mHz). In the Pc5 range (~1-7 mHz), where a distinct frequency is evident at ~5 mHz in ground H, the satellite showed significantly shifted frequency peak at ~3 mHz. Also, the power is found to be nearly two orders higher at satellite compared to ground. The band-pass filtered time series in the Pc5 range appeared to be distinctly different at CHAMP and ground (Figure 6b) with cross correlation (CC) <0.2. The signatures at satellite are found to be dominated by the EEJ profile at ~1527-1531 UT with amplitude 10 times higher than ground Pc5. Therefore, the Pc5 pulsations at CHAMP are severely modified by the dominant latitudinal profile of EEJ currents.

In the Pc4 band (7-22 mHz), ground H component showed frequencies at ~11 and 17 mHz, whereas satellite compressional component showed frequency peaks at ~7, 12 and 17 mHz. Thus, additional frequency at 7 mHz is noticed at satellite, which was not present in the ground observations. The frequency peak at ~11 mHz at ground seems to be shifted to ~12 mHz at satellite, whereas the frequency peak at 17 mHZ is evident at both ground and satellite. The time series band pass filtered in Pc4 range (Figure 6c) showed poor match (CC ~0.2) between the oscillations at satellite and ground. In order to eliminate the contribution from 7 mHz frequency peak at satellite, we applied a narrow band pass filter within Pc4 band in frequency regime 9-22 mHz (Figure 6d). Though improved (CC ~0.4), the oscillations at satellite and ground still appeared to be different, even after removing the additional frequency contribution. But on applying a further narrow band pass filter between 14-22 mHz (Figure 6e), which removes the frequencies <14 mHz from the satellite observations, a good match (CC ~0.8 at 15s lag) between satellite and ground is observed. This indicates that the observations of geomagnetic pulsations having frequencies <15 mHz at LEO satellite are highly contaminated. This imposes limitations on the detection of Pc4-5 oscillations at CHAMP satellite during daytime.

## 5. Observations

In this section, we delineate the characteristics of the observed frequencies by studying its LT, longitudinal and seasonal variation of the PSD (not normalised) in each frequency bin. Here it should be noted that the LT considered in this study is the magnetic local time at equatorial crossing for a given satellite pass.

### 5.1 Local time variation

Figures 7 and 8 show the LT variation of the running average of PSD of frequency peaks from six bins. In order to avoid the regional effects, we have confined the study to two longitudinal sectors viz., Indian (Geographic Longitude: 70°-90°E) (Figure 7) and American (Geographic Longitude: 260°-290°E) (Figure 8). In two years of time, CHAMP spans all LTs nearly 6 times. The average values are computed by taking a running average of PSD for a data window of 2h, shifted by 0.5h in successive steps (solid curve in Figures 7 and 8). The error bars (vertical lines) represent the standard error. Here, one can notice that in both the longitudinal sectors, a clear LT dependence is seen in the frequency bins 1 and 2 which corresponds to a latitudinal wavelength of 7°-22° (Figures 7(a, b) and 8(a, b)), with a maximum power near noon that reduces to much smaller values at night. Also a secondary peak is evident at ~15 h in Bin 1, in the Indian sector. In American sector, the secondary peak is not well defined and appeared early in LT at ~14 h. In bins 3-4 (latitudinal wavelength 4°-7°), although less significant, an enhancement towards noon is even evident, over Indian (Figures 7 (c, d)) and American (Figures 8(c, d)) sectors. Whereas, bins 5-6, (Figures 7 (e, f), 8(e, f)) did not show any systematic LT dependence in both the longitudinal sectors and also a considerably less power is apparent in these bins compared to that in bins 1-4. The maximum PSD values from bins 5-6 are less than the minimum values of power from bins 1-4. The average power is observed to decrease from bin 1-6 in both sectors.

## 5.2 Longitudinal variation

Figure 9 shows the longitudinal variation of the average PSD in each frequency bin. Here, the LT is fixed to 10-13h when the ionospheric conductivity and hence the daytime ionospheric currents are stronger. The average values are estimated by taking the running average of PSD for a longitudinal window of 30°, stepping forward by 10°, with three-points smoothing. The average variation of the PSD in the frequency bins 1, 2 and 3 (Figures 9a, 9b and 9c) show four maxima i.e., around 0°, 90°, 200° and 280°E longitudes. Frequency peaks from Bin4 show a three peak patterns with maxima located near 90°, 190° and 270°E longitudes. Here the peak at 0° seems to be absent. The longitudinal variation in Bin5 and Bin6 appears almost flat. Nevertheless, few isolated peaks are observed in these frequency bins, but their amplitudes are very small and they are not found to be consistent with those in bins 1-4. Therefore, we conclude that no systematic zonal pattern is present for the higher frequency bins that correspond to smaller latitudinal wavelengths of 3°-4°. At the same time, the latitudinal wavelengths between 4°-22° show well defined four peaks in power, located at ~0°, 90°, 200° and 280°E longitudes.

## 5.3 Seasonal variation

In order to study the seasonal dependence of the PSD during 10-13 LT, we have used February, March and April months to represent vernal equinox; May, June and July months for June solstice; and November, December and January months for December solstice. During 2008-09, near September equinox, the CHAMP was almost always outside the 10-13 LT range and hence we could not study the variation for the autumnal equinox, thereby confining equinoxial studies to Spring months only. It can be noticed from Figure 10 that the average PSD value from Bin1 is maximum during equinox in both sectors (Red bars in Figure 10a and 10b). Also clear equinoxial maximum is evident for Bin2 in the American sector, while in the Indian sector, equinoxial value is slightly higher than December solstice, confirming equinoxial high (dark blue bars). Bin3 and Bin4 show maxima at June and December solstices respectively, in both the longitudinal sectors. Bin5 and Bin6 do not show any significant seasonal dependence.

## 6. Discussion and Summary

The present analysis notice the existence of frequencies <20 mHz in all CHAMP passes during quiet geomagnetic conditions in the total (compressional) magnetic field component. These frequencies observed by LEO satellite can be of temporal or spatial origin. The temporal fluctuations (also known as geomagnetic pulsations) having frequency >15 mHz are well studied using LEO satellite data, however lower frequency pulsations mainly constituting Pc4-5 and Pi2 bands (< 15 mHz) faced problems in its detection at LEO satellites [*Balasis et al.,* 2013, *Thomas et al.,* 2015]. A few researchers have noticed the presence of background frequencies <15 mHz in the compressional component at CHAMP satellite [*Heilig et. al.,* 2007 and *Thomas et al.,* 2015] which they thought to be non-temporal oscillations, but attributed to different origins such as latitudinal (spatial) structures of ionospheric currents and crustal anomalies. Hence the source of these frequencies is not well understood yet. Similar to *Thomas et al.,* [2015] who noticed contamination of Pi2s, in this paper we have shown that these lower background frequencies (power ~$10^2$-$10^3$ $nT^2$/Hz) can severely alter the pulsations of Pc4-5 band (typical power ~10 $nT^2$/Hz). Therefore, study of these lower frequencies present at LEO satellites is of great importance. The present work aims to delineate the characteristics of the frequencies <20 mHz identified in the CHAMP magnetic field variations, which can assist in determining the origin of these low frequencies.

Firstly, we examined whether the observed frequencies are of temporal origin. As many ULF Pc3-5 pulsations are caused due to changes in the solar wind parameters, we checked the solar wind conditions and geomagnetic activity, during the days considered in the present study. Low$\sum$Kp values together with low level of Np, SW velocity and IMF Bz make sure the low solar activity. In addition, we examined the dependence of power of the observed frequencies on the magnitude of SW velocity and also on the cone angle. It is found that the average power in each frequency bin is independent of SW velocity (Figure 3) and cone angle (Figure 4), suggesting the observed frequencies in the present study are not Pc oscillations caused by the solar wind or geomagnetic activity. We also ensured the presence of the observed lower frequencies in the CHAMP data, even during the period of negligible Np (< 1), and hence confirm that these are not Pc3-4 oscillations [*Heilig et. al.,* 2010]. The other geomagnetic pulsations such as Pi2 can also fall in the band of frequencies reported here. But the lower geomagnetic activity with IMF Bz nearly zero discard the existence of Pi2 oscillations as well.

Thus it is established that the observed frequencies in the present study are not of temporal origin. If the reported frequencies are due to the spatial structures of the quiet time ionospheric phenomenon [*Thomas et al.,* 2015], then the observed frequencies are expected to have characteristics similar to that of ionospherec current systems. Therefore, the present paper examines the characteristics of these frequencies such as LT, longitudinal and seasonal dependence.

A clear LT dependence with maximum power during noon and minimum at night is evident in Bin1,2,3,4. The power is observed to be nearly 2-3 orders higher during noon compared to night (Bin1-3). This day-night difference in the power may suggest that the daytime phenomenon (such as ionospheric dynamo driven currents) is playing a significant role in the occurrence of these frequencies. Whereas, no clear LT variation is observed in frequency Bin5 and 6.

Further, the longitudinal variation near local noon hours (10-13 LT) shows four peak structure for the frequencies from Bin 1,2,3, with maxima at 0°, 90°, 200° and 280°E longitudes. Also to some extent, the four peak pattern is present in Bin4. This average longitudinal pattern matches well with the earlier reports of longitudinal variation of EEJ based on LEO satellite missions [*Jadhav et. al.* 2002; *Alken and Maus* 2007*; Luhr et.al.*2008], which they attributed to the non-migrating tidal activity, in particular to DE3 tidal mode [*Luhr et.al.,* 2008]. The secondary peak observed near 15 LT in Figure 7 and 8 may be associated with counter electrojet (CEJ). The prominence of the secondary peak is more in the Indian sector compared to the American sector. This may be in accordance with *Vichare and Rajaram* [2011] who noticed that CEJs are more prone to occur over the Indian region than American longitudes.

The seasonal variations of the spectral powers over Indian and American sectors show an equinoctial maximum for Bin 1,2. These observations are in broad agreement with the previous reports about the seasonal variations of EEJ using various LEO missions [*Jadhav et. al.,* 2002; *Alken and Maus* 2007*; Luhr et.al.,* 2008] and ground based observations [*Campbell* 1997]. It can be noted that the latitudinal profile of the residual magnetic field presented in Figure 1a clearly marks the signature of EEJ along with other variations. The first two to three frequency peaks of the observed residual field (Figure 1b) match very well with those of the modelled EEJ (Figure 1c). Thus, the present work demonstrates that the reported frequencies <10 mHz (Bin 1-2) mainly correspond to the latitudinal profile of the EEJ. The LT, longitudinal and seasonal characteristics of Bin 1-2 also suggest its association with EEJ currents, thereby confirming the contribution of EEJ in the reported frequencies <10 mHz. It is also possible that, other current systems can have similar frequencies corresponding to the Bins 1-2. However, EEJ being the strongest among various currents flowing in the equator to mid latitude region, the frequencies will be dominated by the EEJ currents.

The seasonal variations of the spectral powers show solstice maximum for Bin 3,4 in both Indian and American sectors. The solstice maximum for Bin 3,4 is consistent with the seasonal variation of Sq at mid-to-low latitudes [*Campbell* 1997, *Vichare et al.,* 2012]. Moreover, the asymmetry in the Sq current system is normally enhanced during solstice months [*Yamashita and Iyemori* 2002]. Ionospheric dynamo driven by tidal as well as gravity waves can produce the Sq magnetic field variations of frequencies corresponding to Bin 3-4 (wavelength: 4-7°).

Thus, the characteristics of the frequencies from Bin1-4 (<15 mHz) seem to indicate that the dayside ionospheric E-region currents are responsible for their occurrence. *Heilig et al.,* [2007] had suggested that these lower frequencies (<10 mHz) could be due to crustal anomalies those are not properly removed by POMME model. This possibility is not valid, as this consequence should not be LT and seasonal dependent. However the contribution of the spread F associated magnetic fluctuations having spatial scales ~100 Km noticed by *Luhr et. al.,* [2002] during pre midnight hours cannot be neglected.

Therefore, the explanation for the observed lower frequencies (<15 mHz) can be as follows. As polar LEO satellite traverses over different quiet time ionospheric current systems, it monitors the spatial distribution of these currents over different latitudes. The spatial structures associated with different current systems can cause certain frequencies in the geomagnetic field variations recorded by CHAMP (as described in Section 2). As the shape and size of the spatial structures of the ionospheric currents undergo strong day to day variability due to the influence of neutral winds, the frequency of the spectral peak associated with them also vary. The frequencies consistently observed in the present study can be therefore thought of as the latitudinal structures of various ionospheric currents monitored by CHAMP during its transit. For frequencies <15 mHz, which correspond to a latitudinal wavelength of 4°-22° (Table 1), the present study demonstrates the characteristics of the E-region ionospheric currents and therefore supports the proposed scenario. Whereas for frequencies >15 mHz (latitudinal wavelength <4°) a clear dependence on LT, longitude and season could not be identified, and hence may not be associated with the ionospheric dynamo action. Therefore, our work indicates that the spatial structures of daytime ionospheric currents can produce frequencies <15 mHz (wavelengths >4°) in the LEO satellite observations. These frequencies can be observed only by the polar LEO satellite, due to its motion from pole to pole and hence inherent to LEO satellite observations.

Therefore, present paper strongly recommends that the consideration of these intense and intrinsic frequencies is very vital while studying relatively smaller amplitude geomagnetic pulsations (Pi2, Pc4-5) using polar LEO measurements, which otherwise may lead to the misinterpretation. Considering the tremendous data available through previous and ongoing LEO satellite missions, it is of paramount importance to take a notice of this aspect while studying pulsations using LEO satellites.

**Acknowledgement**

Authors are grateful to Information System and Data Center (ISDC) for providing high quality 1s magnetic field data from CHAMP satellite. The data set used is CH-ME-2-FGM-NEC and is available on request from ISDC at http://isdc.gfz-potsdam.de/. We also thank USGS (http://cdaweb.gsfc.nasa.gov/cgi-bin/eval1.cgi) for making available 1s data from the ground observatory, Boulder. We are deeply in debt to Prof. R. Rajaram for all the valuable discussions.

**Figure Captions:**

Figure 1: (a) Typical latitudinal profile of residual total (compressional) magnetic field (red curve) observed during a daytime CHAMP pass on 01 December 2008 over 239°E longitude at ~09:40 LT and the EEJ profile obtained by fitting empirical model of EEJ (blue curve). Log plot of power spectrum for (b) CHAMP pass and (c) modelled EEJ profile, together with 99% confidence level shown by dotted curves.

Figure 2: Histograms of (a) $\sum$Kp index and (b-f) the average SW and IMF parameters during the considered geomagnetic quiet days of the years 2008-2009.

Figure 3: Variation of the average power of frequency peaks with SW velocity for (a) frequency Bins 1-3 and (b) Bins 4-6. Error bars indicate the standard errors.

Figure 4: Variation of the average power of frequency peaks with cone angle for (a) frequency Bins 1-3 and (b) Bins 4-6. Error bars indicate the standard errors.

Figure 5: Log plot of PSD of the total magnetic field recorded at a daytime CHAMP pass between ±50° magnetic latitude during 0225-0250 UT on 04 March 2008 with Np <1 cm$^{-3}$. Red arrows mark the frequency peaks identified and magenta curve denote the 99% confidence level.

Figure 6: Pc4-5 event on 01 August 2008 between 15:15-1535 UT. (a) PSD of the CHAMP compressional (blue solid curve) and BOU H component (black dotted). Time series of satellite compressional and BOU H component filtered (b) in Pc5 band (1.5-6.5 mHz), (c) in Pc4 band (7-22 mHz), (d) narrow band passed (9-22 mHz) and (e) narrow band passed (14-22 mHz). Magnetic latitude of CHAMP is given at the top of Figures 6b and 6d. The universal time (UT) and magnetic local time (MLT) values are given at the bottom of Figure 6c and 6e. The scale for the satellite component is shown on the left side and that for ground H is shown on the right side.

Figure 7: LT dependence of frequency peaks from (a) Bin1, (b) Bin2, (c) Bin3, (d) Bin4, (e) Bin5 and (f) Bin6, with error bars, over Indian sector.

Figure 8: Same as Figure 7, but for American sector.

Figure 9: Longitudinal variation of frequency peaks from (a) Bin1, (b) Bin2, (c) Bin3, (d) Bin4, (e) Bin5 and (f) Bin6, with error bars, during 10-13 LT.

Figure 10: Average values of PSD during Equinox, June and December solstices for satellite passes between 10-13 LT at (a) Indian and (b) American Sectors. Values for bins 1-2, bins 3-4 and bins 5-6 are shown in top, middle and bottom panels respectively.

**Table 1:** Frequency bins and corresponding latitudinal wavelengths

| Bin | Frequency range (mHz) | Latitudinal wavelengths (deg) | Km |
|---|---|---|---|
| **Bin1** | 3 – 6 | 22 – 11 | 2540-1270 |
| **Bin2** | 6 – 9 | 11 – 7 | 1270-850 |
| **Bin3** | 9 – 12 | 7 – 5 | 850-635 |
| **Bin4** | 12 – 15 | 5 – 4 | 635-510 |
| **Bin5** | 15 – 18 | 4 – 3.6 | 510-425 |
| **Bin6** | 18 – 21 | 3.6 – 3 | 425-365 |

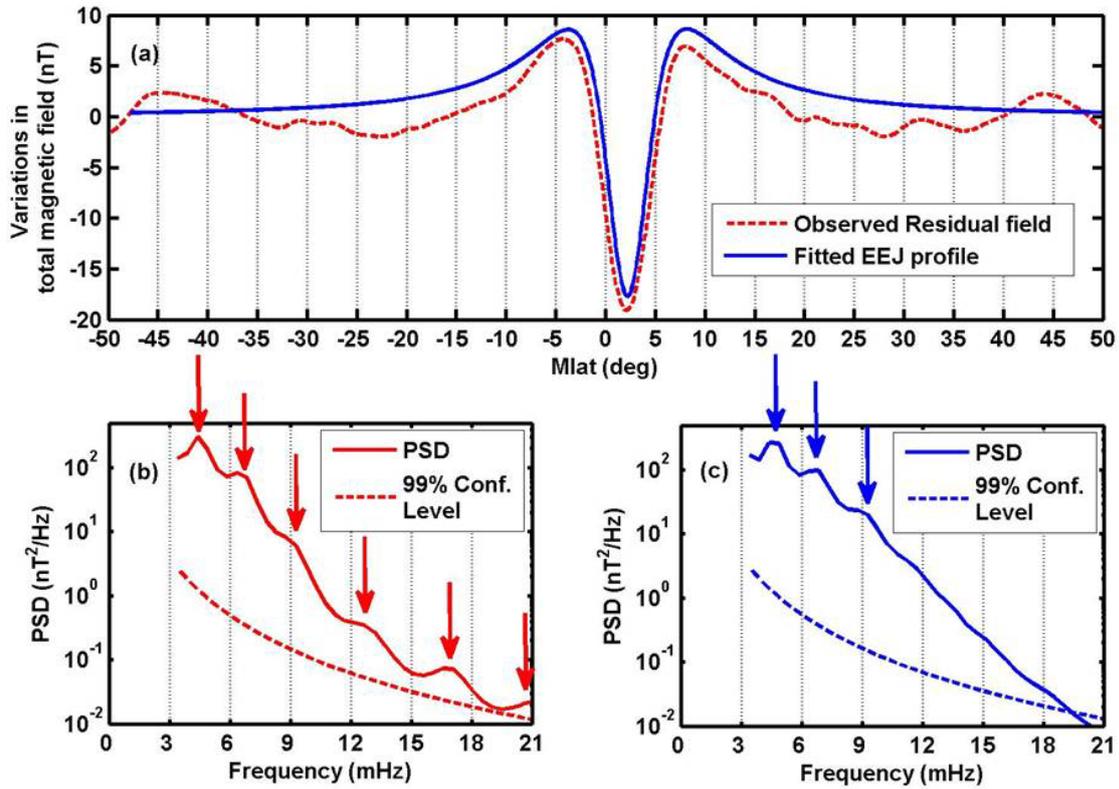

Figure 1: (a) Typical latitudinal profile of residual total (compressional) magnetic field (red curve) observed during a daytime CHAMP pass on 01 December 2008 over 239°E longitude at ~09:40 LT and the EEJ profile obtained by fitting empirical model of EEJ (blue curve). Log plot of power spectrum for (b) CHAMP pass and (c) modelled EEJ profile, together with 99% confidence level shown by dotted curves.

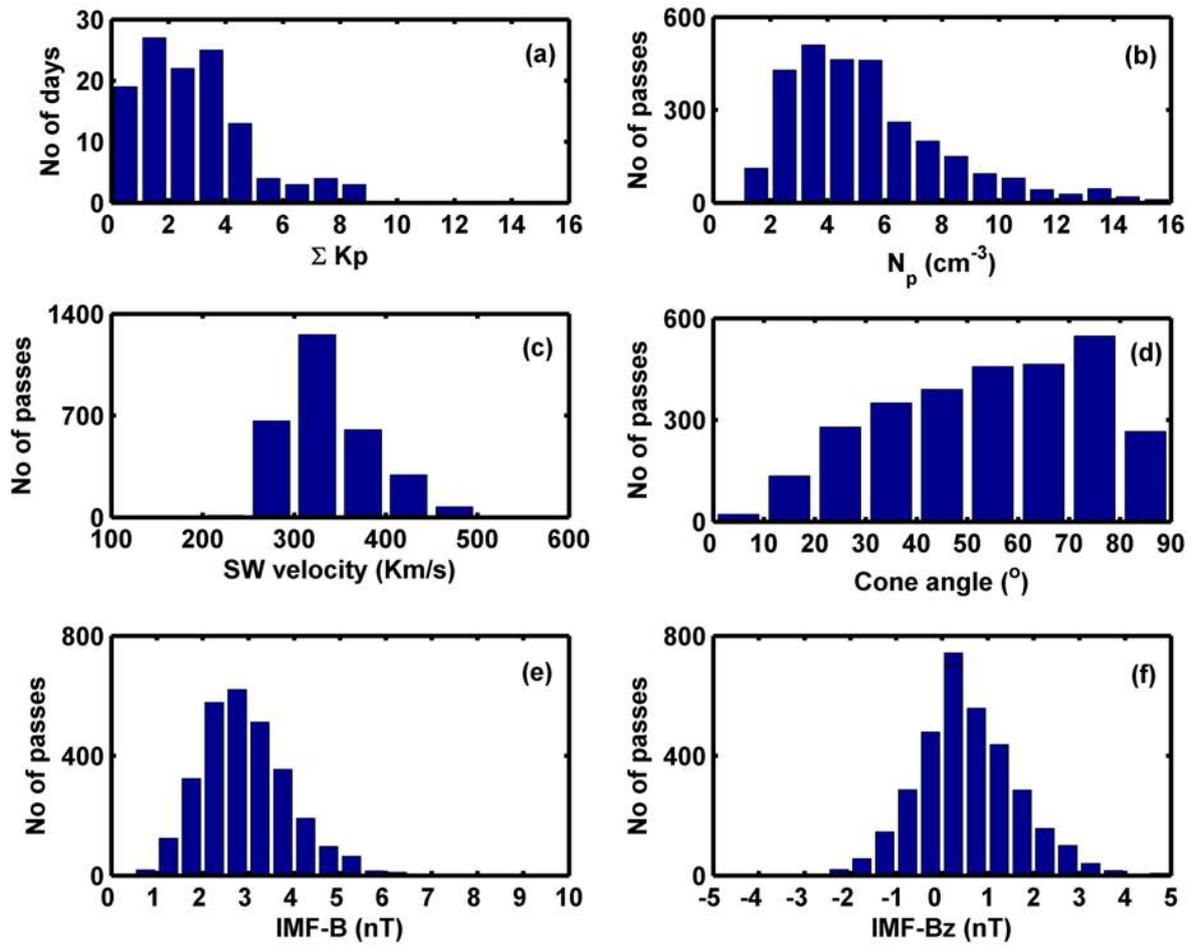

Figure 2: Histograms of (a) ∑Kp index and (b-f) the average SW and IMF parameters during the considered geomagnetic quiet days of the years 2008-2009.

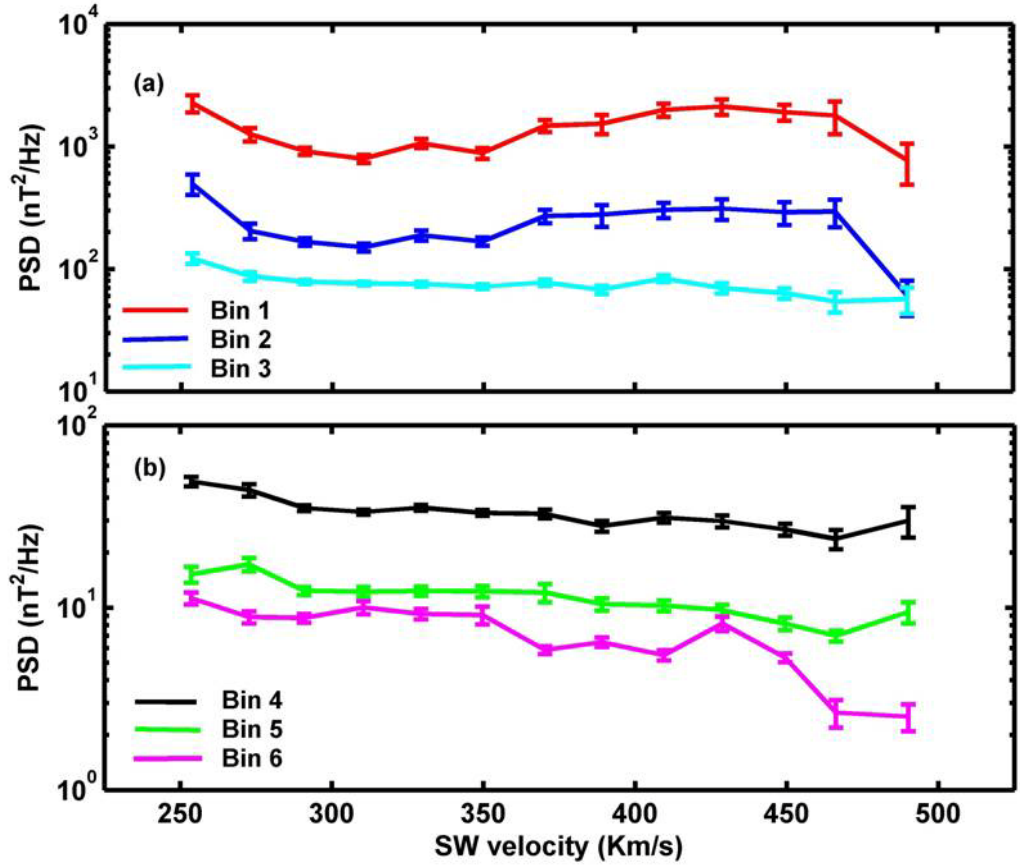

Figure 3: Variation of the average power of frequency peaks with SW velocity for (a) frequency Bins 1-3 and (b) Bins 4-6. Error bars indicate the standard errors.

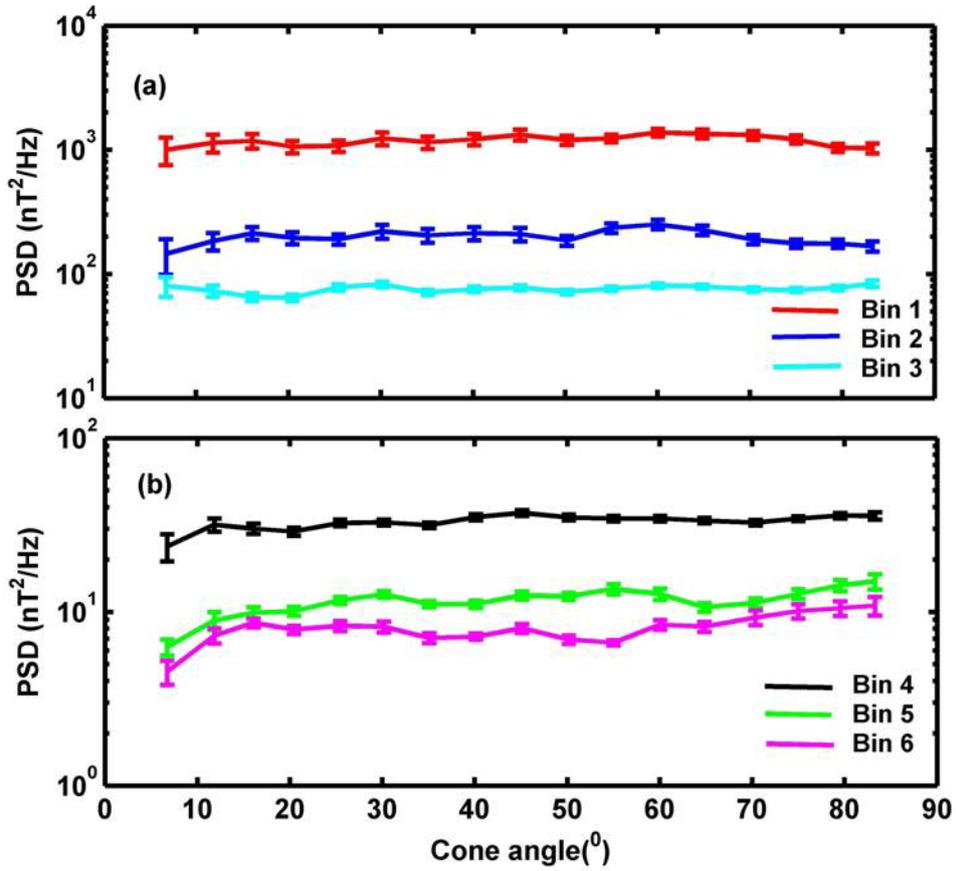

Figure 4: Variation of the average power of frequency peaks with cone angle for (a) frequency Bins 1-3 and (b) Bins 4-6. Error bars indicate the standard errors.

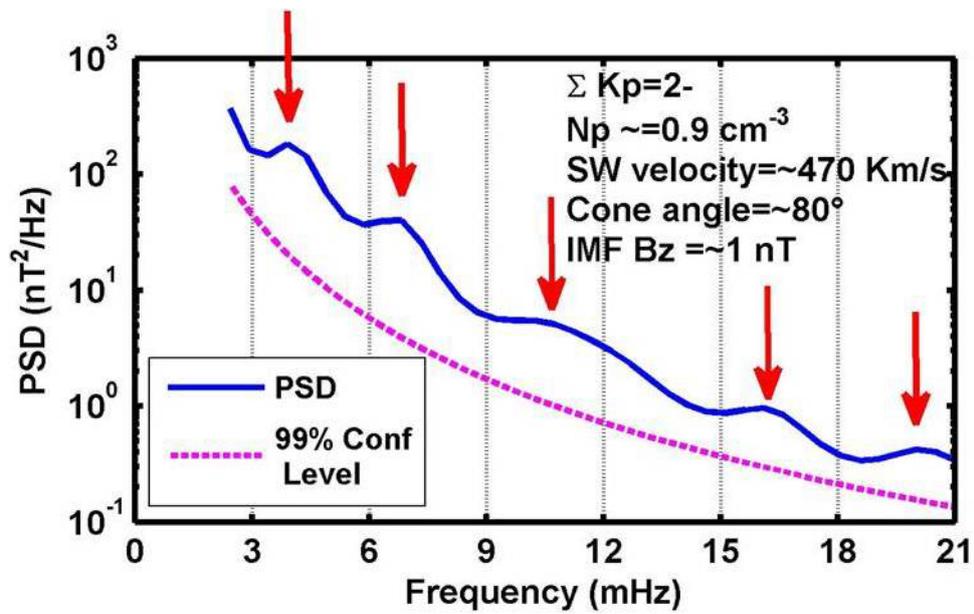

Figure 5: Log plot of PSD of the total magnetic field recorded at a daytime CHAMP pass between ±50° magnetic latitude during 0225-0250 UT on 04 March 2008 with Np <1 cm$^{-3}$. Red arrows mark the frequency peaks identified and magenta curve denote the 99% confidence level.

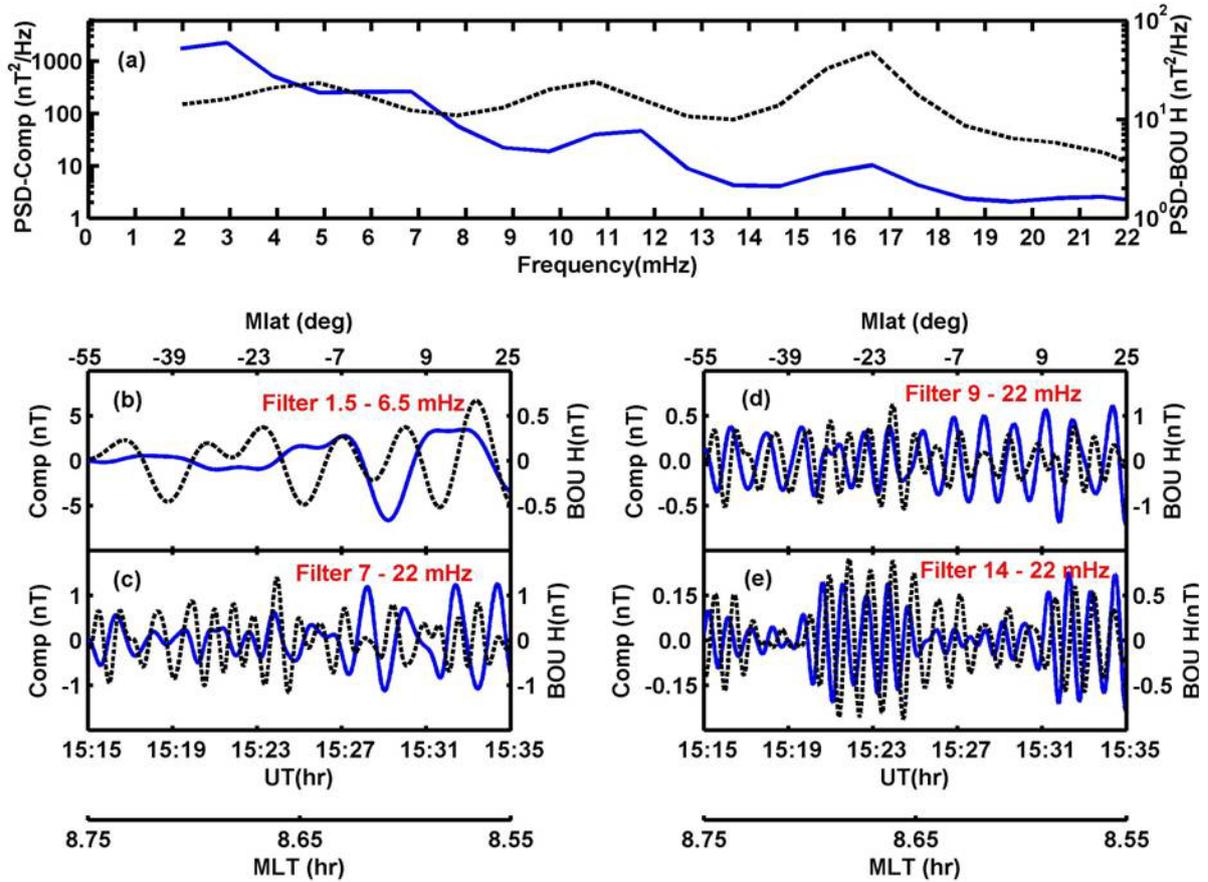

Figure 6: Pc4-5 event on 01 August 2008 between 15:15-1535 UT. (a) PSD of the CHAMP compressional (blue solid curve) and BOU H component (black dotted). Time series of satellite compressional and BOU H component filtered (b) in Pc5 band (1.5-6.5 mHz), (c) in Pc4 band (7-22 mHz), (d) narrow band passed (9-22 mHz) and (e) narrow band passed (14-22 mHz). Magnetic latitude of CHAMP is given at the top of Figures 6b and 6d. The universal time (UT) and magnetic local time (MLT) values are given at the bottom of Figure 6c and 6e. The scale for the satellite component is shown on the left side and that for ground H is shown on the right side.

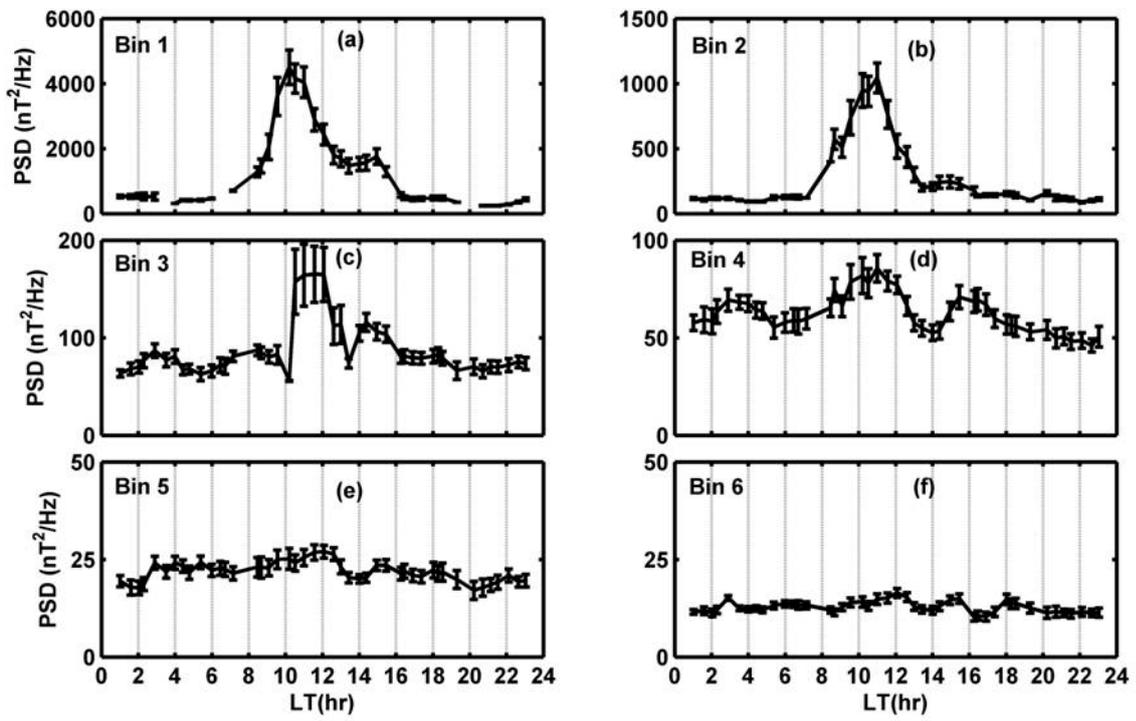

Figure 7: LT dependence of frequency peaks from (a) Bin1, (b) Bin2, (c) Bin3, (d) Bin4, (e) Bin5 and (f) Bin6, with error bars, over Indian sector.

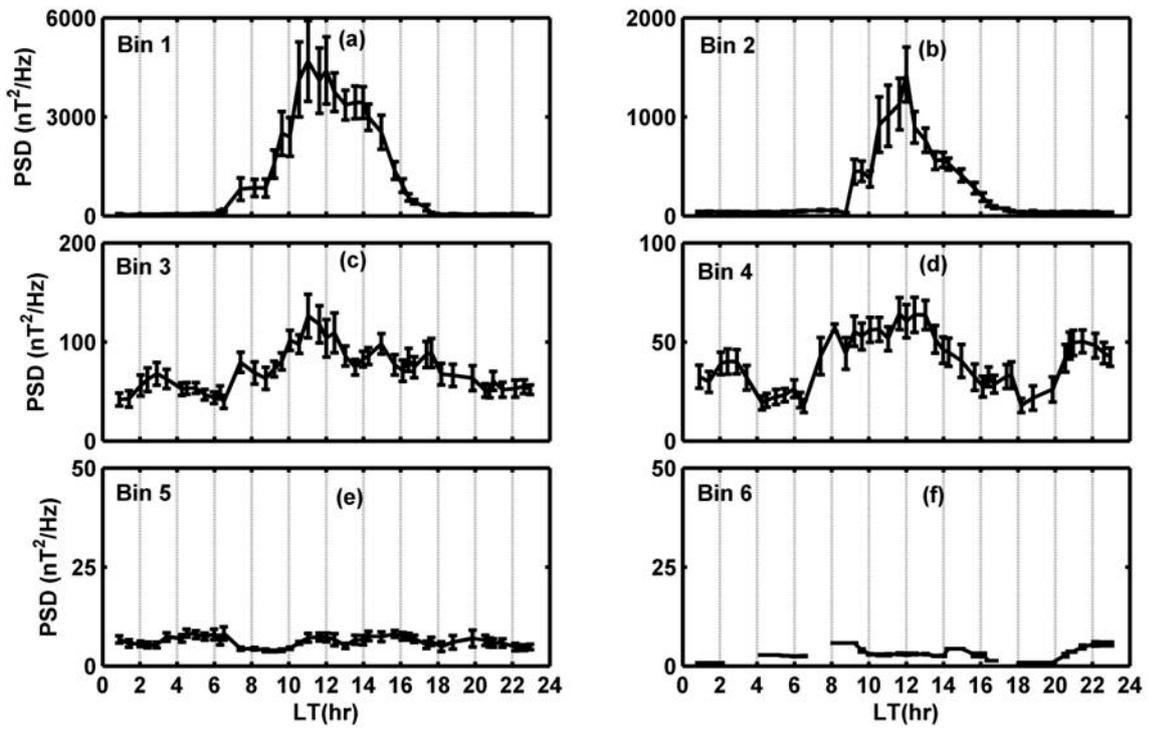

Figure 8: Same as Figure 7, but for American sector.

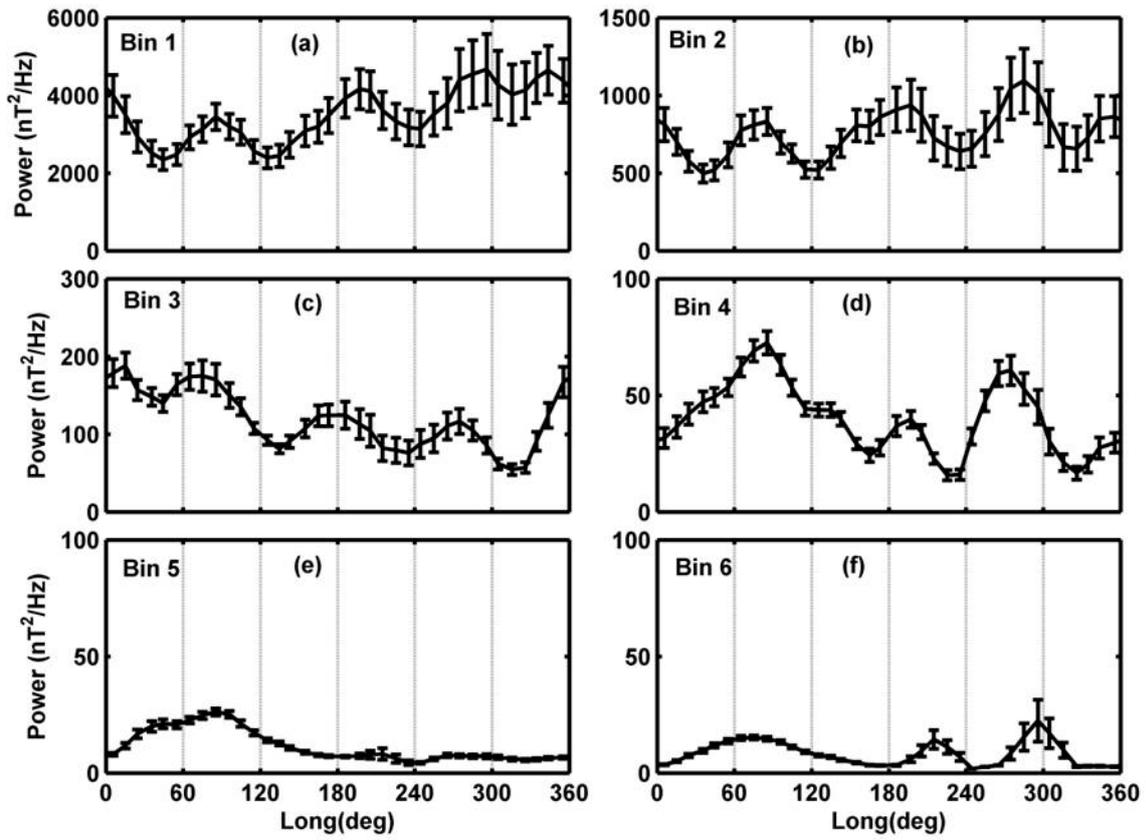

Figure 9: Longitudinal variation of frequency peaks from (a) Bin1, (b) Bin2, (c) Bin3, (d) Bin4, (e) Bin5 and (f) Bin6, with error bars, during 10-13 LT.

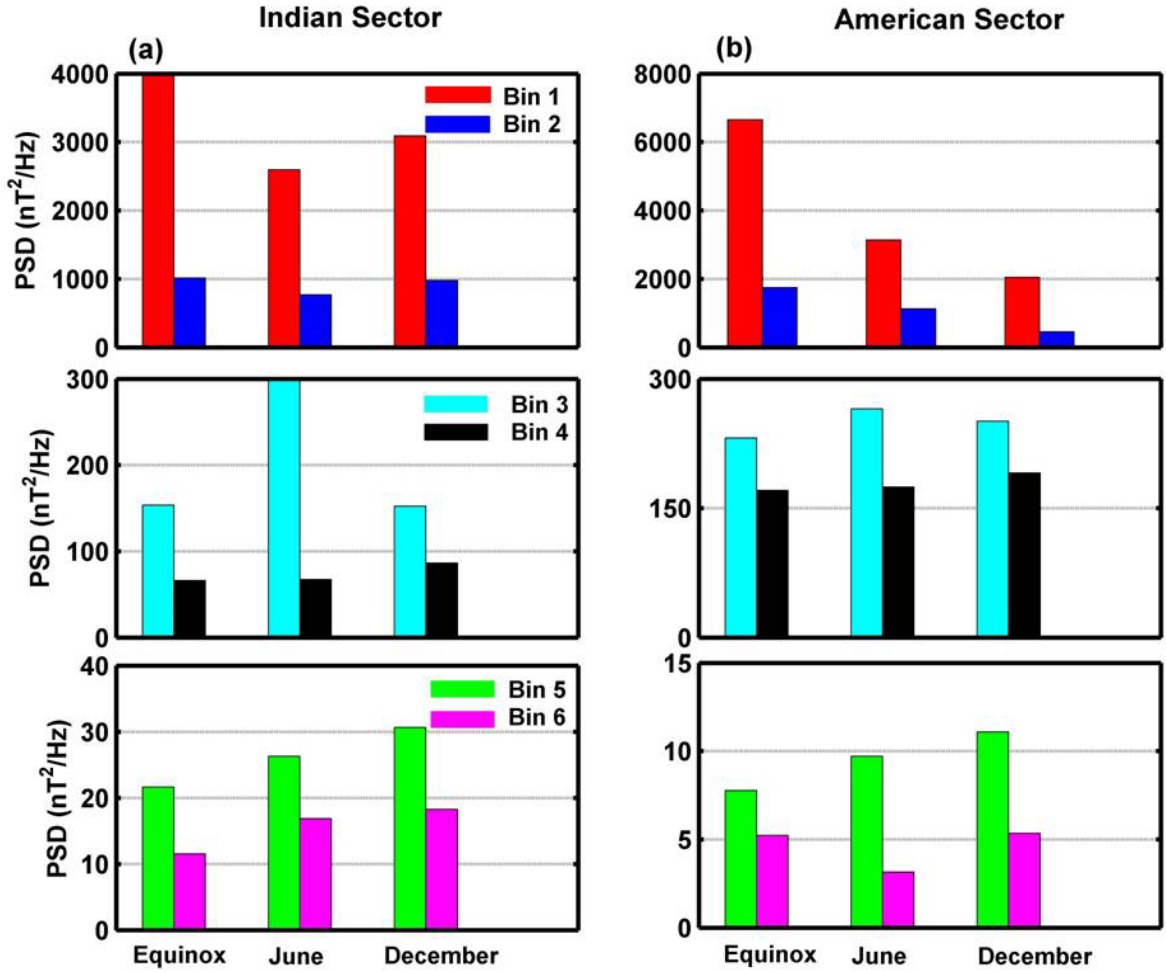

Figure 10: Average values of PSD during Equinox, June and December solstices for satellite passes between 10-13 LT at (a) Indian and (b) American Sectors. Values for bins 1-2, bins 3-4 and bins 5-6 are shown in top, middle and bottom panels respectively.